\documentstyle[aps,multicol]{revtex}

\newcommand{\ppp}{^{\raisebox{0.5mm}{\tiny +}}}

\draft
\title{Phase transitions in the antiferromagnetic XY model with a
       kagom\'{e}  lattice}
\author{S. E. Korshunov}
\address{L.D.Landau Institute for Theoretical Physics,
Kosygina 2, Moscow 117940, RUSSIA}
\date{June 22, 2001}

\begin{document}
\maketitle

\begin{abstract}
The ground state of the antiferromagnetic $XY$ model with
a {\em kagom\'{e}} lattice is known to be characterized by a well
developed accidental degeneracy.
As a consequence the phase transition in
this system consists in unbinding of pairs of fractional vortices.
Addition of the next-to-nearest neighbor (NNN) interaction leads
to stabilization of the long-range order in chirality (or staggered
chirality). We show that the phase transition, related with
destruction of this long-range order, can happen as a separate
phase transition below the temperature of the fractional vortex
pairs unbinding only if the NNN coupling is extremely weak, and
find how the temperature of this transition depends on coupling
constants. We also demonstrate that the antiferromagnetic ordering
of chiralities and, accordingly, the presence of the second phase
transition are induced by the free energy of spin wave
fluctuations even in absence of the NNN coupling.
\end{abstract}
\pacs{PACS numbers: 75.10.Hk, 64.60.Cn, 74.80.-g}

  \begin{multicols}{2}
\section{Introduction}

The antiferromagnetic $XY$ model can be defined by the Hamiltonian
\begin{equation}
H=J_1\sum_{NN}\cos(\varphi_{i}-\varphi_{j}),        \label{NN}
\end{equation}
where $J_1>0$ is the coupling constant,
$\varphi_{i}$ describes the orientation of the classical planar spin
\makebox{${\bf s}_i=(\cos\varphi_{i},\;\sin\varphi_{i})$},
belonging to the site $i$ of some regular lattice,
and the summuation is performed over the pairs of nearest neighbors
(NN) on this lattice.
The ground state of such model on a {\em kagom\'{e}} lattice (Fig. 1)
is known to have a huge accidental (that is not related to
the symmetry of the Hamiltonian) degeneracy \cite{HR}.

For $J_1>0$ the minimum of the energy of each triangular plaquette is
achieved when the three spins belonging to it form the angles
$\pm 2\pi/3$ with each other. In addition to the possibility
of a simultaneous rotation of all three spins such
arrangement is also characterized by the two-fold discrete degeneracy.
When on going clockwise around the plaquette the spins rotate
clockwise (anticlockwise), the plaquette can be ascribed the positive
(negative) chirality $\sigma =\pm 1$.
In the ground state of the antiferromagnetic  $XY$ model with
triangular lattice the plaquettes with positive and negative
chiralities regularly alternate with each other \cite{MS}.

In any ground state on a {\em kagom\'{e}} lattice
the variables  $\varphi_{i}$ analogously acquire only three values
which differ from each other by $2\pi/3$.
However, the requirements for the arrangement of
chiralities are less rigid than on triangular lattice and accordingly
the ground state in addition to the continuous $U(1)$ degeneracy
(related with an arbitrary simultaneous rotation of all spins)
is characterized by a well developed discrete degeneracy
(of the same type as in the 3-state antiferromagnetic Potts model)
leading to a finite residual entropy per site \cite{HR,Els}.

The accidental degeneracy persists if the interaction function
in Eq. (\ref{NN}) differs from the pure cosine (remaining even),
but is removed by the
presence of interactions with more distant neighbors.
In particular, for the ferromagnetic sign ($J_2<0$) of
the next-to-nearest neighbors (NNN) interaction the minimum of
\begin{equation}
H_{2}=J_1\sum_{NN}\cos(\varphi_{i}-\varphi_{j})
 +J_2\sum_{NNN}\cos(\varphi_{i}-\varphi_{j})          \label{NNN}
\end{equation}
is achieved in one of the so-called $\sqrt{3}\times\sqrt{3}$
states \cite{HKB}
with a regular alternation of positive and negative chiralities.
An example of such state is shown in Fig. 2(a).
Here and further on we use the letters A, B and C to denote three
values of $\varphi_{i}$ which differ from each other by $\pm 2\pi/3$
(for definiteness let us assume $\varphi_{\rm B}=\varphi_{\rm A}+2\pi/3$,
$\varphi_{\rm C}=\varphi_{\rm A}+4\pi/3$).
This state has the same structure as the ground state of a planar
antiferromagnet with triangular lattice (or, to put it more
accurately, can be obtained by the natural truncation of it).
On the other hand the NNN interaction of the opposite sign
($J_2>0$) favors the ferromagnetic arrangements of chiralities, which
is achieved in the so-called ${\bf q}=0$ states \cite{HKB},
see Fig. 2(b).

For both signs of the NNN interaction the degeneracy of the ground
state is reduced to $U(1)\times Z_2$ \cite{GB}.
This suggests the possibility of the two phase transitions,
one of which is associated with unbinding of vortex pairs
[the Berezinskii-Kosterlitz-Thouless (BKT) transition
\cite{Ber1,Ber2,KT,Kost}] and the other with proliferation of the
Ising type domain walls.
The number of the systems with the same degeneracy of the ground state
includes, in particular, the antiferromagnetic $XY$ model with
triangular lattice \cite{MS,KU} and the fully frustrated $XY$ model
with square lattice \cite{Vill,Hals}.

For the case of weak NNN interactin ($|J_2|\ll J_1$) the energy of
a domain wall (per unit length) is proportional to $|J_2|$ and
the logarithmical interaction of vortices to $J_1$.
If domain walls and vortices were completely independent
one would then expect the phase transition related with
breaking of the discrete symmetry group to take place at lower
temperature than the BKT transition.

In the present work we analyse the mutual influence between different
classes of topological excitations in the antiferromagnetic $XY$
model with a {\em kagom\'{e}} lattice and weak NNN interaction
and demostrate that the phase transition related with the domain walls
proliferation
(i) can happen as a separate phase transition only for extremely weak
NNN interaction, (ii) is not of the Ising type and (iii) the temperature
of this transition is {\em not} proportional to the strength of
the NNN interaction as one could naively expect.
We also show that at very low temperatures the free energy of spin
waves leads to stabilization of the antiferromagnetic ordering
of chiralities even in absence of the NNN interaction.

The results can be of interest in relation with possible presence
of weak easy-plane anisotropy in Heisenberg {\em kagom\'{e}}
antiferromagnets \cite{RCC}, which is indirectly confirmed by
recent investigations of susceptibility \cite{WDV} in
(H$_3$O)Fe$_3$(SO$_4$)$_2$(OH)$_6$ .
The other class of physical systems,
which allows for experimental realization of the considered model,
consists of Josephson junction arrays \cite{ML} and
two-dimensional superconducting wire networks \cite{HXB}
in perpendicular magnetic field.
In such systems the role of $\varphi_{i}$ is played by the phase of
the superconducting order parameter, and equivalence with the
antiferromagnetic $XY$ model is achieved when the magnitude of the
magnetic field corresponds to half-integer number of
superconducting flux quanta per each triangular plaquette.
A superconducting array, which can be described by the same model
in absence of the external magnetic field, can be constructed
with the help of so-called $\pi$-junctions \cite{PJ}.

\section{Zero temperature: the equivalent solid-on-solid model}

It has been already mentioned in the Introduction that the set
of the ground states of the Hamiltonian (\ref{NN}) is equivalent
(up to a simultaneous rotation of all spins)
to that of the 3-state antiferromagnetic Potts model \cite{HR,Els}.
Any triangular plaquette of a {\em kagom\'{e}} lattice has to
contain some permutation of the three values of $\varphi_{i}$
($\varphi_{\rm A}$, $\varphi_{\rm B}$ and $\varphi_{\rm C}$), which
can be identified with the three states ($\alpha=1,2,3$) of the
antiferromagnetic Potts model.

The degeneracy of such set of ground states
can be discussed in terms of the zero-energy domain walls.
If one consideres a $\sqrt{3}\times\sqrt{3}$ ground state [Fig. 2(a)],
it turnes out possible to construct the state with the same energy
by permutation (for example) of the form
$\varphi_{\rm B}\Longleftrightarrow\varphi_{\rm C}$ inside any closed
loop formed by the sites with \mbox{$\varphi_{i}=\varphi_{\rm A}$},
{\em etc}.
Such closed loop [a simplest example of which is shown in Fig. 3(a)]
can be considered as the zero-energy domain wall separating two
different $\sqrt{3}\times\sqrt{3}$ states with the opposite
signs of staggered chirality.
Any domain wall with zero energy is formed by elementary links
which have to join each other at the angles of $\pm 2\pi/3$
[Fig. 3(b)].
Each link separates two triangular plaquettes with the same
chirality, that is with the opposite signs of staggered chirality.
The states with infinite (unclosed) domain walls are also possible.

There exists \cite{HeR} the exact mapping of the set of the ground
states of the 3-state antiferromagnetic Potts model
onto the states of the solid-on-solid (SOS)
model in which the "height" variables ${\bf u}({\bf r})$ are defined
on the sites ${\bf r}$ situated at the centres of hexagonal
plaquettes of a {\em kagom\'{e}} lattice.
These cites are shown in Fig. 1 by empty circles.
They form the triangular lattice we shall denote $\cal T$.
Each site of the {\em kagom\'{e}} lattice
can be associated  with some bond ${\bf rr}'$ of $\cal T$ and each
variable $\varphi_{i}=\varphi_{\rm A},\varphi_{\rm B},\varphi_{\rm C}$
with the Potts variable $\alpha({\bf rr'})\equiv\alpha({\bf r'r})$
defined on this bond.

Since each triangular plaquette of a {\em kagom\'{e}} lattice
should always contain three different variables $\varphi_{\rm A}$,
$\varphi_{\rm B}$ and $\varphi_{\rm C}$, %[corresponding to
% three different Potts model variables $\alpha$ ($\alpha=1,2,3$)],
one can associate them with three basic vectors ${\bf a}_\alpha$
(${\bf a}_{1}+{\bf a}_2+{\bf a}_3=0$, see Fig. 4) of some auxilary
triangular lattice, which we shall denote ${\cal T}_a$
($a=|{\bf a}_\alpha|$ being its lattice constant).
The height variables ${\bf u}({\bf r})$,  which acquire
the values ${\bf u}\in{\cal T}_a$, can be then introduced
following the rule
\begin{equation}
{\bf u}({\bf r}')=\left\{\begin{array}{ll}
{\bf u}({\bf r})+{\bf a}_{\alpha({\bf rr}')}
& \mbox{for } {\bf r}'={\bf r}+{\bf e}_\alpha \\
{\bf u}({\bf r})-{\bf a}_{\alpha({\bf rr}')}
& \mbox{for } {\bf r}'={\bf r}-{\bf e}_\alpha,
\end{array}\right.                                     \label{3}
\end{equation}
where ${\bf e}_\alpha$ are the three basic vectors of ${\cal T}$
(${\bf e}_1+{\bf e}_2+{\bf e}_3=0$),
as soon as the value of ${\bf u}({\bf r}_0)$ is chosen for an
arbitrary site ${\bf r}_0$ \cite{HeR}.

This defines the correspondence between the states of the
antiferromagnetic Potts model and of the "vector" SOS model, in which
the height variables  ${\bf u}({\bf r})\in {\cal T}_a$
have to satisfy the constraint
\begin{equation}
|{\bf u}({\bf r})-{\bf u}({\bf r}')|=a               \label{UU}
\end{equation}
on all pairs of neighboring sites of ${\cal T}$.
By using the known properties of the exact solution \cite{B}
of the 3-state antiferromagnetic Potts model with external field
coupled to staggered chirality Huse and Rutenberg \cite{HR} have
demonstrated (and also have confirmed this conclusion by numerical
calculation) that such vector SOS model,
in the partition function of which all allowed configurations of
heights are counted with the same weight,
is situated exactly at the point of the roughening transition,
where (for $|{\bf r}_1-{\bf r}_2|\gg 1$)
\begin{equation}
\langle[{\bf u}({\bf r}_1)-{\bf u}({\bf r}_2)]^2\rangle\approx
\frac{3a^2}{\pi^2}\ln|{\bf u}({\bf r}_1)-{\bf u}({\bf r}_2)|.
                                                     \label{ln}
\end{equation}
Therefore any additional perturbation supressing the fluctuations
will lead to transition of the system into the flat phase,
in which the fluctuations of ${\bf u}$ are convergent.

According to constraint (\ref{UU})
the variables ${\bf u}({\bf r})$ on neighboring sites have to be
different from each other, so that even the most flat state is
formed by the regular alternation of three different values of
${\bf u}$. The transition into the flat phase can be more
transparently discussed in terms of the variables
\begin{equation}
 {\bf n}({\bf R})\equiv\frac{{\bf u}({{\bf r}})+{\bf u}({{\bf r}'})
+{\bf u}({{\bf r}''})}{3}
\end{equation}
describing the average height at each of the
plaquettes of $\cal T$.
The variables ${\bf n}({{\bf R}})$ are defined at the sites ${\bf R}$
of the honeycomb lattice $\cal H$, which is dual to $\cal T$, and
acquire the values ${\bf n}({\bf R})\in{\cal H}_a$, where ${\cal H}_a$
is the honeycomb lattice which is dual to ${\cal T}_a$ (Fig. 4).
In terms of the original spin variables the flat states
[in which all variables ${\bf n}({\bf R})$ are equal to each other]
correspond to $\sqrt{3}\times\sqrt{3}$ states, and zero-energy
steps, the presence of which leads to their roughening,
to the zero-energy domain walls separating different
$\sqrt{3}\times\sqrt{3}$ states.

The large scale properties of the vector SOS model introduced
above (and of its further generalizations) can be analysed with the
help of the multi-component sine-Gordon model with the same symmetry.
The (dimensionless) Hamiltonian of such sine-Gordon model can be
chosen in the form
\begin{equation}
H_{SG}=\int_{}^{}d^2{\bf R}\left\{\frac{KQ^2}{2}
      [\nabla{\bf n}({\bf R})]^2+
y\sum_{\alpha=1}^{3}\cos[{\bf Q}_\alpha{\bf n}({\bf R})]\right\}.
                                                          \label{SG}
\end{equation}
The first term in Eq. (\ref{SG}) describes the effective stiffness
(of entropic origin)
which can be associated with the fluctuations of ${\bf n}({\bf R})$,
whereas the second term
favors the values of ${\bf n}({\bf R})$ which belong to ${\cal H}_a$.
Here ${\bf Q}_\alpha$ are the three basic vectors
of the triangular lattice reciprocal to $T_a$, so
\[{\bf Q}_\alpha^2=\frac{16\pi^2}{3a^2};~~~
  {\bf Q}_1+{\bf Q}_2+{\bf Q}_3=0. \]

Analogous Hamiltonian (with the opposite sign of the second term)
and equivalent vector Coulomb gas have been investigated by Nelson
in relation with dislocation mediated melting in two-dimensional
crystals \cite{N}.
Alternatively the Hamiltonian of the form (\ref{SG}) can be
interpreted as a simplified model for pinning of a two-dimensional
crystal by a periodic substrate (cf. with Ref. \onlinecite{NH}).
Note, however, that in contrast to real two-dimensional crystals,
the accurate description of which requires to distinguish between
compression and shear moduli,
in our system the displacement ${\bf n}$ takes place in some auxilary
space (and not in the real space) and, therefore, only one elastic
modulus can be introduced.

The renormalization group equations of Ref. \cite{N},
describing the evolution of $K$ and $y$ with the change
of the length scale $L$, in our notation can be rewritten as
\begin{eqnarray}
\frac{dK}{dl} & = & \frac{3\pi}{8}y^2             \eqnum{8a}     \\
\frac{dy}{dl} & = & \left(2-\frac{1}{4\pi K}\right)y-\pi y^2,
                                                  \eqnum{8b}
\addtocounter{equation}{1}
\end{eqnarray}
where $l=\ln L$.
The corresponding flow diagram is schematically shown in Fig. 5,
where $K_c\equiv 1/(8\pi)$. It suggests that the roughening transition
takes place when the renormalized value of the effective stiffness
$K$ is equal to $K_c$.
The vector SOS model described above (which is known to be at the
point of its roughening transition \cite{HR}) can therefore be
associated with some point belonging to the left separatrix.

\section{Phase transition(s) associated with vortex pairs
unbinding}

At finite temperature T other types of fluctuations (requiring
finite energy) become possible, in particular formation of vortex
pairs.
Vortices are point-like topological excitations (the local minima of
the Hamiltonian), the existence of which is related with the
multivaluedness of the field $\varphi$. On going around each vortex
$\varphi$ experiences a continuous twist which adds to $\pm 2\pi$.
At low temperatures all vortices are bound in neutral pairs by their
logarithmical interaction \cite{Ber2}.

With increase in temperature this interaction becomes renormalized due
to mutual influence of different vortex pairs and
becomes screened at the temperature $T_{\rm BKT}$
of the BKT transition \cite{Ber2,KT,Kost},
which leads to dissociation of vortex pairs and exponential decay of
correlations of $\exp(i\varphi)$
(in contrast to algebraic decay \cite{Ber1} at $T<T_{\rm BKT}$).
The value of the helicity modulus $\Gamma$,
describing the effective stiffness of spin system with respect to
infinitely small twist, at the temperature of vortex pairs
dissociation is known to satisfy the universal relation \cite{NK}:
\begin{equation}
T_{\rm BKT}=\frac{\pi}{2}\Gamma(T_{\rm BKT}).          \label{BKT}
\end{equation}

Huse and Rutenberg \cite{HR} have argued that since at $T=0$
the antiferromagnetic $XY$ model with {\em kagom\'{e}} lattice
is characterized by the long range order in $\exp(i3\varphi)$
rather than in $\exp(i\varphi)$, the phase transition in this system
should consist in unbinding of pairs of fractional vortices with topological
charges $\pm 1/3$ and not of the ordinary (integer) vortices.
The strength of the logarithmical interaction of fractional vortices
is decreased by the factor of 9 in comparison with that of integer
vortices \cite{RCC}, therefore relation (\ref{BKT}) should be replaced
by \begin{equation}
T_{\rm FV}=\frac{\pi}{18}\Gamma(T_{\rm FV}),             \label{GFV}
\end{equation}
where $T_{\rm FV}$ is the temperature of the phase transition,
associated with unbinding of pairs of fractional vortices.

The value of $\Gamma$ in any ground state of the antiferomagnetic
$XY$ model with {\em kagom\'{e}} lattice
is equal to $\Gamma_0=(\sqrt{3}/4)J_1$.
Substitution of $\Gamma_0$ into Eq. (\ref{GFV})
[instead of $\Gamma(T_{\rm FV})$] allows to obtain for $T_{\rm FV}$
the estimate (from above) of the form
\begin{equation}
T_{\rm FV}\approx\frac{\pi\sqrt{3}}{72}J_1\approx0.075 J_1, \label{TFV}
\end{equation}
which turns out to be in reasonable agreement with
the results of numerical simulations by Rzchowski \cite{R},
who by using two different criteria has found
$T_{\rm FV}\approx 0.070$ and $T_{\rm FV}\approx 0.076$.
In Ref. \cite{CKP} the same estimate for $T_{\rm FV}$ has been obtained in
a less straightforward way with the help of the duality transformation
\cite{JKKN,Knops}.

The fractional vortices cannot exist by themselves
(in absence of domain walls).
A fractional vortex appears at every point where elementary links
forming a domain wall meet each other at a wrong angle
($\pi/3$ or $\pi$ instead of $2\pi/3$).
The same happens in the antiferromagnetic $XY$ model with triangular
lattice \cite{KU},
the ground state of which also has $\sqrt{3}\times\sqrt{3}$ structure.
Fig. 6(a) shows an example of a domain wall containing one such
special point.
It separates the domain wall into two segments, one of which is formed
by the sites with $\varphi_{i}=\varphi_{\rm C}$ and the other by the
sites with $\varphi_{i}=\varphi_{\rm B}$.

When crossing the first segment the state to the right of the wall
should be obtained from the state to the left by the permutation of
A and B, whereas for the second segment the state to the right
should be obtained by the permutation of A and C.
This introduces the descrepancy of $2\pi/3$ which can be localized
on a semi-infinite line terminating at the special point
(for example on the line X-Y-Z).
In order to locally minimize the energy X, Y and Z should be replaced
by A, B and C respectively when going from above, and by B, C and A
when going from below.
The misfit of $2\pi/3$ has to be compensated by a continuous
twist of $\varphi$, which is equivalent to the fractional vortex with
the topological charge $-1/3$.

In terms of the vector SOS model each fractional vortex corresponds
to the point on going around which the height variable ${\bf n}$
changes by $\Delta {\bf n}$ with $|\Delta {\bf n}|=a$.
That means that at each fractional vortex a step with the height
$\Delta {\bf n}$ (or, to put it more presicely, a set of steps with
the total height $\Delta {\bf n}$) terminates or begins.
Accordingly the fluctuations of the SOS model provide
the additional contribution to the interaction of the fractional
vortices related with the difference in entropy between the
configurations with different positions of step termination points.
At the point of roughening transition of the SOS model, as well as in
the rough phase, this additional interaction (which can be expressed
in terms of the correlation function of the dual $XY$ model
\cite{Sw}) is also logarithmic.
Its presence shifts $T_{\rm FV}$ upwards and diminishes
the mutual influence between fractional and integer vortices.
It is known \cite{LGT} that such mechanism in principle can lead
to appearance of the separate phase transition, associated with
unbinding of pairs of integer vortices, at temperatures above
$T_{\rm FV}$.

On the other hand at finite temperatures the equivalence with
the SOS model is no longer exact.
One has to remember that the whole multitude of what we describe
as flat states of the vector SOS model in terms of the original spin
variables corresponds (for given $\varphi_{\rm A}$) to only six
different
$\sqrt{3}\times\sqrt{3}$ states, which can be obtained from the state
shown in Fig. 2(a) by all possible permutations of A, B and C.
In Fig. 4 the sites of ${\cal H}_a$, which correspond to equivalent
states in terms of $\varphi_{i}$, are designated by the same numbers.
In the zero-temperature partition function of the vector SOS model
the properties of the zero-energy domain walls separting such states
(in particular, two closed loops formed by such walls cannot cross
each other but can be situated inside each other or touch each other)
allow to count them as different states of the SOS model
\cite{HeR,vB}.
At finite temperature it becomes possible for a set of steps separating
two physically equivalent states to terminate at the point where all
these steps merge together \cite{Lyu}.
The energy $E_D$ of such defect is finite and proportinal to $J_1$:
$E_D=c_D J_1$, where $c_D$ is of the order of one.

In terms of the multi-component sine-Gordon model (\ref{SG})
such defects correspond to dislocations
of the field ${\bf n}$, the (elementary) Burgers vectors of which
${\bf b}_\alpha$ ($\alpha=1,2,3$) are given,
as can be seen from Fig. 4, by
\begin{equation}
{\bf b}_1={\bf a}_3-{\bf a}_2,~~
{\bf b}_2={\bf a}_1-{\bf a}_3,~~
{\bf b}_3={\bf a}_2-{\bf a}_1.                           \label{BV}
\end{equation}
An example of a dislocation is schematically shown in Fig. 6(b).
It is formed by a neutral pair of fractional vortices that are
sitting on two domain walls which cannot be transformed into
a single domain wall.
The letter X denotes the site on which
$\varphi_{i}\approx(\varphi_{\rm A}+\varphi_{\rm B})/2$.
In the vicinity of this site the values of $\varphi_{i}$ slightly
deviate from those implied by the letters A, B and C.
Successive application of the rule (\ref{3}) along the perimeter of
any closed loop surrounding point X sums up to
$\Delta{\bf n}={\bf a}_2-{\bf a}_1$.

The renormalization of the dislocation fugacity $z=\exp(-c_D J_1/T)$
with the change of the length scale can be described \cite{NH} by
\begin{equation}
\frac{dz}{dl} = \lambda_z z+2\pi z^2                    \label{RG3}
\end{equation}
where
\begin{equation}
\lambda_z=2-\frac{KQ^2b^2}{4\pi}\equiv 2-4\pi K         \label{LaZ}
\end{equation}
In the vicinity of the roughening transition ($K\approx K_c$)
the exponent $\lambda_z$, describing the renormalization of $z$, is
close to $\lambda_z^0=3/2$, which corresponds to the fast growth of
$z$. Comparison of $\lambda_z$ with $\lambda_y=2-1/(4\pi K)$
shows that $y$ and $z$ are never simultaneously irrelevant.
In that respect the situation is quite analogous to what is
encountered when considering the conventional (ferromagnetic) $XY$
model with weak but relevant (low-order) anisotropy \cite{JKKN,PU}.

The presence of dislocations (or, to put it more accurately,
of dislocation pairs) leads also to appearance in the right-hand
side of Eq. (8a) of the additional (negative) term proportional
to $z^2$.
The presence of this term shifts the flow (see Fig. 5) from the
separatrix to the area which corresponds to the rough phase of
the SOS model.
% In the close vicinity of $T_{\rm FV}$ it would be also necessary to
% include into Eq. (\ref{RG3}) the terms which explicitely
% take into account that dislocation is a complex object,
% formed by a pair of fractional vortices,
% but for lower temperatures this is not necessary.

On the other hand
the unrestricted growth of $z$ under the renormalisation means
that the system will contain the finite concentration of free
dislocations, which transforms the rough phase of the SOS model
into the disordered phase of the six-state model.
The decay of correlations in this phase can be characterized by
a finite correlation radius $\xi_z$, which can be found as  the
length-scale at which $z_{\rm R}$ (the renormalized value of $z$)
becomes of the order of one.
$\xi_z$ defines the scale at which the additional (entropic)
interaction of the fractional vortices induced by the fluctuation
of the domain walls is screened. The finiteness of $\xi_z$
closes even the hypothetical possibility for dissociation of
pairs of integer vortices to take place as a separate phase
transition at $T>T_{\rm FV}$.

\section{The case of the ferromagnetic NNN interaction}

Inclusion into consideration of the interaction with more distant
neighbors leads to removal of the accidental degeneracy and stabilizes
the states with either ferromagnetic or antiferromagnetic ordering
of chiralities of triangular plaquettes.

In the case of the ferromagnetic NNN interaction \mbox{($J_2<0$)}
the energy is minimal in  one of the $\sqrt{3}\times\sqrt{3}$ states
with uniform staggered chirality [Fig. 2(a)].
Fig. 3  shows two examples of a domain wall separating two
different $\sqrt{3}\times\sqrt{3}$ states with opposite signs of
staggered chirality. In the case of only NN interaction such
domain wall (consisting of elementary links making angles of
$2\pi/3$ with each other) simply costs no energy.
The presence of a weak ferrromagnetic NNN interaction
makes the energy of such domain wall
(per elementary link) $E_{\rm DW}$ equal to $-3J_2>0$.
Here and further on we are interested only in the case $|J_2|\ll J_1$,
when the values of the variables $\varphi_{i}$ remain close to those
shown in Fig. 3 not only away from the wall, but also in the
vicinity of the wall.

Note that such wall can fluctuate (make turnes, form kinks, etc.)
without having to pay the energy proportional to $J_1$
and, therefore, naively one could expect that the temperature of
the phase transition, related with proliferation of such
walls and leading to destruction of the long-range order in
staggered chirality, should be determined entirely by $J_2$.
Such conclusion, implicitely based on the comparison of the energy of
an infinite domain wall with the negative entropic contribution to its
free energy (the Peierls argument \cite{Pei}),
does not take into account that the presence of an
infinite domain wall leads also to supression of the entropy (because
it decreases the possibilities for formation of closed domain wall
loops) and in some cases does not work.

In Sec. II we have discussed the properties of the vector SOS model
to which the antiferromagnetic $XY$-model with {\em kagom\'{e}}
lattice is equivalent at $T=0$.
In the partition function of this model
all allowed configurations of heights are counted with the same weight.
In the case of the analogous model with a finite positive energy
of a step (which corresponds to $J_2<0$) the same partition function
is reproduced in the limit of $T\rightarrow \infty$.
For any small but finite ratio $-J_2/T>0$ the SOS model is shifted
from the point of roughening transition into the ordered (flat) phase.
On the other hand we also know that at finite temperatures the
possibility of dislocation creartion tends to shift the same system
into the disordered phase.
One has to consider the competition of these two effects.

In the vicinity of $K=K_c$ the renormalization equations (8)
can be rewritten as
\begin{eqnarray}
\frac{dX}{dl} & = & {Y}^2          \eqnum{15a}            \\
\frac{dY}{dl} & = & X{Y}+\alpha{Y}^2 ,
                                         \eqnum{15b}
\end{eqnarray} \addtocounter{equation}{1}
where
\begin{equation}
X=2\left(1-\frac{K_c}{K}\right),~~{Y}=\frac{1}{\pi \alpha},~~
\alpha=-\frac{1}{\sqrt{6}} .
\end{equation}
The solution of Eqs. (15) for arbitrary $\alpha$ allows one to find
that the critical behavior of the correlation
radius $\xi$ in the vicinity of the transition is given \cite{Y} by
\begin{equation}
\ln\xi\propto\left(\frac{K_c}{\Delta K}\right)^{\overline{\nu}} ,
                                                    \label{Xi}
\end{equation} where $\Delta K$ is the deviation from the
phase transition. In our problem for $E_{\rm DW}\ll T$ the ratio
$\Delta K/K_c$ is proportional to $E_{\rm DW}/T$.

The case of $\alpha=0$ corresponds to Kosterlitz renormalization
group equations \cite{Kost} for the standard BKT transition, which
give $\overline{\nu}=1/2$.
The case of $\alpha=+1/\sqrt{6}$ has been considered by Nelson
\cite{N}, who has found $\overline{\nu}=2/5$.
The solution of the same equations for $\alpha=-1/\sqrt{6}$ gives
$\overline{\nu}=3/5$.

If the fugacity of dislocations $z$ is so small that even when growing
under renormalization it remains much smaller than one up to $L\sim\xi$
[where the renormalization following Eqs. (15) stops anyway and
fluctuations of ${\bf n}$ are frozen],
the system remains in the ordered (flat) phase of the SOS model,
that is in the phase with long range order in staggered chirality.
On the other hand if $z_R$ (the renormalized
value of $z$) manages to become of the order of one when
the renormalized value of $y$ is still small,
the system finds itself in the disordered phase.
The estimate for the temperature $T_{\rm DW}$ of the phase transition
separating these two regimes (and associated with the proliferation of
domain walls) can be obtained from the relation
$z_R(\xi)\sim 1$, which is equivalent to
\begin{equation}
\ln(1/z)\sim\lambda^0_z\ln\xi.
\end{equation}

In our problem
\begin{equation}
\ln(1/z)= c_D J_1/T
\end{equation}
and
\begin{equation}
\ln\xi\sim\left(\frac{T}{E_{\rm DW}}\right)^{\overline{\nu}},
                                                     \label{LNX2}
\end{equation}
which leads to
\begin{equation}
T_{\rm DW}\sim\left(\frac{E_{\rm DW}}{J_1}\right)^
{\frac{\overline{\nu}}{1+\overline{\nu}}}J_1
                        \propto J_2^{3/8}J_1^{5/8} , \label{TDW1}
\end{equation}
Away from the critical region the behaviour of $\xi$ can be
described with the help of the self-consistent harmonic approximation
\cite{Sait}, which gives $\overline{\nu}=\overline{\nu}_0=1$.
That means that with increase of $J_2/J_1$ the dependence (\ref{TDW1})
is replaced by $T_{\rm DW}\propto(J_2 J_1)^{1/2}$.

Note that the analysis which has led to Eq. (\ref{TDW1})
has been based on the assumption that all fractional vortices are
bound in pairs, and, accordingly, is valid only for
$T_{\rm DW}<T_{\rm FV}$.
On the other hand the pairs of fractional vortices cannot dissociate
at temperatures lower than $T_{\rm DW}$, because for $T<T_{\rm DW}$ the
fractional vortices in addition to their logarithmical
interaction are bound also by domain walls (with a finite free energy
per unit length) which connect them with each other.

Therefore the two available possibilities are $T_{\rm DW}<T_{\rm FV}$
and a single phase transition, whereas the scenario with
$T_{\rm DW}>T_{\rm FV}$ is impossible.
Analogous conclusions have been earlier achieved in relation
with hypothetical unbinding of fractional vortices in planar
antiferromagnet with triangular lattice \cite{KU}.
It is hardly surprising that the same conclusions are valid for
the system, the ground state of which is practically
identical to that of the antiferromagnetic $XY$ model with triangular
lattice, the only difference being that one quarter of sites is
absent.

The proliferation of the low energy domain walls
[of the type shown in Fig. 3(b)]
leads to intermixing of six different states [which can be obtained
from the state shown in Fig. 2(a) by arbitrary permutation of A, B,
and C] and therefore should not be expected to be of the Ising type.
Note that the domain walls are possible only between the states with
the different signs of staggered chirality.
The six-state model with analogous statistics of domain walls can be
defined \cite{KU} by the partition function
\begin{equation}
Z_{\rm 6st}=\left(\prod_{{\bf R}}\sum_{t_{\bf R}=1}^6\right)
            \prod_{NN}W(t_{\bf R}-t_{\bf R'})
                                                          \label{6ST}
\end{equation}
where
\begin{equation}
W(t)=\left\{\begin{array}{lll}
1 & \mbox{ for } t=0 & \pmod{6} \\
w & \mbox{ for } t=1,3,5 & \pmod{6} \\
0 & \mbox{ for } t=2,4 & \pmod{6} . \end{array}\right. \label{W}
\end{equation}
The last line of Eq. (\ref{W}) implies that the domain wall between
the states with the same parity of $t_{\bf R}$ is impossible.

Application of the duality transformation \cite{Dots} to the partition
function (\ref{6ST}) transforms it into analogous partition function
with $W(t)$ replaced by
\begin{equation}
\widetilde{W}(t)=\left\{\begin{array}{lll}
1+3w   & \mbox{ for } t=0 & \pmod{6} \\
{1}    & \mbox{ for } t=1,2,4,5 & \pmod{6} \\
{1-3w} & \mbox{ for } t=3 & \pmod{6} . \end{array}\right.
\end{equation}
The symmetry of $\widetilde{W}(t)$ corresponds to the so-called
cubic model, the phase transition in the six-state version of which
for $\widetilde{W}(1)>\widetilde{W}(3)$
is known to be of the first order \cite{NRS}.
The phase transition at $T_{\rm DW}$ in our system
(at least when it happens at $T_{\rm DW}<T_{\rm FV}$)
therefore also can be expected to be of the first order.

Comparison of the estimate (\ref{TDW1}) with Eq. (\ref{TFV}) shows
that the fulfillment of the relation $T_{\rm DW}<T_{\rm FV}$ requires
$0<-J_2<J_{{\rm max}}$, where $J_{{\rm max}}$ can be estimated
to be of the order of $10^{-3}J_1$.
For $-J_2>J_{\rm max}$ there should be only one phase transition
in the system, %(happening at $T=T_c$ satisfying $T_{\rm FV}<T_c<T_{\rm DW}$),
at which the proliferation of
domain walls is accompanied by the unbinding of all types of vortices.
A detailed description of how it happens still remains to be
constructed, but when the dissociation of pairs of fractional vortices
is forced by the dissapearance of their linear interaction (mediated
by the domain walls which connect them) at temperatures, for which
their logarithmic interaction is already too weak, one can
expect the value of the helicity modulus at $T_c$ to be nonuniversal:
\begin{equation}
\frac{2}{\pi}<\frac{\Gamma(T_c)}{T_c}<\frac{18}{\pi}      \label{GTC}
\end{equation}
Note that the estimate for $J_{{\rm max}}$ has been found by taking
the estimate (\ref{TDW1}) on its face value, that is without
the unknown numerical coefficient, and therefore should be considered
with great caution.

\section{The case of the antiferromagnetic NNN interaction}

For the antiferromagnetic sign ($J_2>0$) of the NNN interaction
the minimum of the Hamiltonian (\ref{NNN}) is achieved in one of
the states with the ferromagnetic ordering of chiralites [Fig. 2(b)].
Such state also allows for construction of a domain wall,
the energy of which (per elementary link) ${E}\ppp_{\rm DW}$ is
proportional to the strength of second neighbor coupling:
$E\ppp_{\rm DW}\approx 3J_2$, see Fig. 7(a).
However, comparison of Fig. 7(a) with Fig. 7(b) shows that
(in contrast to the case considered in Sec. IV) for $J_2>0$ the form
of the state on the other side of the wall is uniquely defined by
the orientation of the wall and is different for different
orientations of the wall.
The descrepancy in $\varphi_{}$ that appears when crossing domain
walls of different orientations should be taken care of by
the fractional vortices which have to appear on {\em all} corners of
domain walls (the same happens in the fully-frustrated $XY$ model
with square lattice \cite{Hals}).
This makes impossible the construction of a closed domain wall the
energy of which is determined entirely by $J_2$ and does not depend
on $J_1$.

For $J_2\ll J_1$ a typical thermally excited defect
(leading to the change of the sign of chirality)
has the  form of a long strip formed by two low energy
domain walls [Fig. 7(c)]. Like in Fig. 6 the
letter X designates the sites with
$\varphi\approx(\varphi_{\rm A}+\varphi_{\rm B})/2$.
In the vicinity of these sites the values of other variables
$\varphi_{i}$ slightly deviate from that shown in the figure.
Analogous strip defects are dominant at low temperatures in the
frustrated $XY$-model with triangular lattice and $f=1/4$ or $f=1/3$
\cite{KVB}.

The energy of such defect is given by $2E_0+2E\ppp_{\rm DW}L$,
where $E_0=c_0 J_1$ ($c_0\approx 0.55$) is the energy of its
termination point and $L$ its length.
For $J_2\ll T\ll J_1$ the average length $\langle L\rangle$ of such
defects is given by the ratio $T/(2E\ppp_{\rm DW})\gg 1$,
whereas their
concentration $c$ is proportional to $\langle L\rangle\exp(-2E_0/T)$.
The relation $c\langle L\rangle^2\sim 1$ defines
the temperature
\begin{equation}
T_*\approx \frac{2}{3}\frac{E_0}{\ln(T_{*}/E\ppp_{\rm DW})}
\end{equation}
above which such defects no longer can be considered as independent.
The same temperature can serve as the (lower bound) estimate for the
temperature $T\ppp_{\rm DW}$ of the phase transition associated with
proliferation of the domain walls and leading to destruction of the
long range order in chirality. For $J_2/J_1\rightarrow 0$
\begin{equation}
T\ppp_{\rm DW}\sim T_*\propto J_1/\ln(J_1/J_2)  \label{TDW2}
\end{equation}
Analogous estimate can be obtained by comparison of the domain wall
energy $E\ppp_{\rm DW}$ with its entropy
\makebox{$S\ppp_{\rm DW}\approx2\exp(-E_K/T)$}
due to possibility of formation of kinks [Fig. 7(d)].
The energy of a kink $E_K$ is very close to $E_0$.
The requirement $F\ppp_{\rm DW}\equiv E\ppp_{\rm DW}-TS\ppp_{\rm DW}=0$
\cite{Pei} gives
\begin{equation}
T\ppp_{\rm DW}\sim E_K/\ln(2T\ppp_{\rm DW}/E\ppp_{\rm DW})
                                                     \label{TDW2'}
\end{equation}
which is again the estimate from below.

Like in the previous case (of the antiferromagnetic ordering of
chiralities) the proliferation of domain walls can take place as
the independent phase transition only at temperatures lower than
$T_{\rm FV}$.
Comparison of Eq. (\ref{TDW2'}) with Eq. (\ref{TFV}) shows that the
fulfillment of relation $T\ppp_{\rm DW}<T_{\rm FV}$ requires
$J_2<{J}\ppp_{\rm max}$, where $J\ppp_{\rm max'}$
can be estimated as $(10^{-4}\div 10^{-5})J_1$.
Also like in the previous case, the proliferation of domain walls is
related with intermixing of six different states and therfore can
hardly be expected to demonstrate the Ising type behavior.

\section{Spin wave fluctuations}

Another  mechanism for removal of accidental degeneracy (which is
traditionally refered to as "ordering due to disorder" \cite{VBCC})
is related with the free energy of continuous fluctuations (spin
waves) \cite{Kaw}.
Expansion of the Hamiltonian (\ref{NN}) up to the second order in
deviations  $\psi_{i}\equiv\varphi_{i}-\varphi_{i}^{(0)}$ of
the variables $\varphi_{i}$ from their values $\varphi_{i}^{(0)}$
in some ground state gives the same answer
\begin{equation}
H^{(2)}=\frac{J_1}{4}\sum_{NN}[-1+(\psi_{i}-\psi_{j})^2]                         \label{HE}
\end{equation}
for all possible ground states, which means that the difference in
the free energy between them can appear only in the second order in
temperature \cite{HR}. That is believed to be not sufficient
for stabilization of a true long-range order related with chiralities.
This conclusion does not take into account the percularities of the
statistical mechanics of the considered system and has to be corrected.

With the help of the numerical calculation (see Appendix)
we have found that the lowest order contribution to the effective
interaction of chiralities of neighboring triangular plaquettes
is of the antiferromagnetic sign
(that is favors $\sqrt{3}\times\sqrt{3}$ state) and corresponds to
\begin{equation}
E^{(0)}_{\rm DW}=\gamma\frac{T^2}{J_1}                    \label{SW}
\end{equation}
where $\gamma\approx 2\cdot 10^{-3}$.
Quantum and thermal anharmonic fluctuations
in Heisenberg {\em kagom\'{e}} antiferromagnet are also
known to favour (at least locally) a planar state with
$\sqrt{3}\times\sqrt{3}$ structure \cite{Chub}.
The same can be said about the fluctuations of the order parameter
amplitude in superconducting wire networks \cite{PH}.

Although $E^{(0)}_{\rm DW}$ defined by Eq. (\ref{SW}) is always much
smaller than the temperature
and in the case of (for example) Ising model would be
insufficient for appearance of the long-range order, in the considered
system the situation is qualitatively different.
Substitution of Eq. (\ref{SW}) into Eq. (\ref{TDW1})
gives a finite value of $T_{\rm DW}$ induced by spin wave fluctuations:
\begin{equation}
T^{(0)}_{\rm DW}\sim \gamma^{3/2}J_1                    \label{TDW0}
\end{equation}
which means that for $\gamma\sim 1$ the long range order in
staggered chirality would survive even up to $T\sim J_1$.
However substitution of the numerically calculated value of $\gamma$
cited above produces an extremely low estimate:
$T^{(0)}_{\rm DW}\sim 10^{-4}J_1$.

Note that the ordering in staggered chirality is noticable only at
length-scales larger than the correlation radius $\xi$.
Substitution of Eq. (\ref{SW}) into Eq. (\ref{LNX2}) shows that for
$T\ll T^{(0)}_{\rm DW}$ the behavior of $\xi(T)$ is given by
$\ln\xi\propto(J_1/\gamma T)^{\overline{\nu}}$.
That means that at $T\rightarrow 0$ there takes place a
continuous reentrant phase transition into the phase without true
long-range order in staggered chirality.

\section{Conclusion}

This work has been devoted to investigation of the phase transitions
in the antiferomagnetic $XY$ model on a {\em kagom\'{e}} lattice
with the special emphasis on accurate consideration of mutual
influence between different classes of topological excitations
(fractional vortices and domain walls).
In particular, we have shown that in the model with only NN
interaction the additional interaction of fractional vortices
related with the entropic contribution from zero-energy domain walls
at finite temperatures becomes short-ranged.
Therefore it can not interfere with the BKT dissociation of
fractional vortex pairs proposed in Refs. \onlinecite{HR} and
\onlinecite{RCC}.

For the case of a finite NNN coupling
(leading to removal of the accidental degeneracy)
we have demonstrated that the  phase transition related with
proliferation of the domain walls can happen as a separate phase
transition below $T_{\rm FV}$
only for very weak NNN interaction, and have found how the
temperature of this transition depends on $J_1$ and $J_2$.
These dependences are essentially different for different signs of
the NNN coupling.
The same results are also applicable for other mechanisms of removal
of the accidental degeneracy, which lead to a finite $E_{\rm DW}$.
Note that
our analysis has been restricted to the case $|J_2|\ll J_1$, so we
have not considered the possibility of the domain wall proliferation
happening above the temperature of the ordinary BKT transition,
associated with appearance of free integer vortices
(like it happens in the case of triangular lattice \cite{KU,LL,new}).

Our conclusions are compatible with the results of
the numerical simulations of Geht and Bondarenko \cite{GB},
who have found (for not too weak NNN interaction, $|J_2|\gtrsim 0.1J_1$)
that the disordering of all degrees of freedom in the
antiferromagnetic $XY$ model with {\em kagom\'{e}} lattice
takes place at the same temperature,
the singularities of the thermodynamic quantities being of the Ising
type.
Recently it has been shown \cite{new} (for the case of triangular
lattice) that when the domain wall proliferation happens as a
continuous phase transition
[at $T/\Gamma(T)>2T_{\rm FV}/\Gamma(T_{\rm FV})$],
the dissociation of pairs of integer vortices has to take place
at $T<T_{\rm DW}$.
Since the same arguments are also applicable for a {\em kagom\'{e}}
antiferromagnet with $J_2<0$, it may be of interest to check
the results of Ref. \cite{GB} with better accuracy.

The long-range order in staggered chirality is favored also by
the spin-wave fluctuations. Our analysis suggests that
the antiferromagnetic $XY$ model with {\em kagom\'{e}} lattice and
only NN interaction presents a unique example of a model
without free parameter in which one of the phase transitions can be
expected
to happen at dimensionless temperature of the order of $10^{-4}$.
Therefore one can conclude that
the numerical simulations of Rzchowskii \cite{R} have
demonstrated no evidence  for selection of a single ground state
down to $T/J_1\approx 10^{-3}$ not because $E_{\rm DW}\propto T^2/J_1$
is not sufficient for that, but simply because the temperature was
not low enough. Comparison with Eq. (\ref{TDW2}) shows that
if the effective interaction of chiralities, induced by the free
energy of spin waves, would be of the opposite sign, the long range
order in chirality would persist up to much higher temperatures.

Experimentally the phase transitions discussed in this work can be
observed in superconducting wire networks or Josephson junction arrays
in the external magnetic field providing one-half of the
superconducting flux quantum $\phi_0=hc/2e$
per each triangular plaquette of a {\em kagom\'{e}} lattice.
In such systems the removal of the accidental degeneracy
is related with the magnetic interaction of the
currents and a finite width of the wires \cite{PH}.
The former of these mechanisms favors the ferromagnetic ordering
of chiralities, whereas for the latter the effect depends on the width
of the wires.

Recent experimental investigation of the aluminum wire network with
{\em kagom\'{e}} structure \cite{HXB} has demonstrated for
$\phi=\phi_0/2$ the presence on the current-volatage curve
of the regions corresponding to different mechanisms of dissipation,
one of which (with an algebraic behavior)
can be associated with unbinding of vortex pairs and the other
with spreading of domains with inversed chiralities \cite{MT}.
The authors of Ref. \cite{HXB} have interpreted this as an evidence
for the presence of two phase transitions.

\section*{Acknowledgments}

This work has been supported in part by the Program
"Statistical Physics" of the Russian Ministry of Science,
by the Program "Scientific Schools of the Russian Federation"
(grant No. 00-15-96747),
by the Swiss National Science Foundation,
and by the Netherlands Organisation for Scientific Research (NWO)
in the framework of Russian-Dutch Cooperation Program.
The author is greatful to A. V. Kashuba for useful discussion
and to I. V. Andronov for assistance in preparation of the figures.

\appendix
\section*{}
The lowest order contribution to the interaction of
chiralities of neighboring triangular plaquettes induced by the spin
wave free energy appears when the partition function
of the  Hamiltonian (\ref{NN}) is expanded up to the second order in
\begin{equation}
H^{(3)}=\frac{J_1}{6}\sum_{NN}[\sin(\varphi_{i}-\varphi_{j})]
        (\psi_{i}-\psi_{j})^3,                      \label{H3}
\end{equation}
and then is averaged with the help of $H^{(2)}$.
The fourth-order term is the same for all the ground
states and therefore of no importance.

The parameter $K_\sigma$ describing the effective interaction of
chiralities of neighboring triangular plaquettes ($a$ and $b$):
\begin{equation}
E(\sigma_a,\sigma_b)=K_\sigma\sigma_a\sigma_b         \label{A2}
\end{equation}
can be then found by calculating the average of
\begin{equation}
V=-\frac{H^{(3)}_a H^{(3)}_b}{T}                      \label{A3}
\end{equation}
where
\begin{equation}
H^{(3)}_a=\frac{\sin\frac{2\pi}{3}}{6}J_1
[(\psi_{1}-\psi_{0})^3+(\psi_{2}-\psi_{1})^3
+(\psi_{0}-\psi_{2})^3]                         \label{A4}
\end{equation}
and expression for $H^{(3)}_b$ can be obtained by replacing in
Eq. (\ref{A4}) $\psi_{1}$ by  $\psi_{3}$ and $\psi_{2}$ by $\psi_{4}$.
The indices from $0$ to $4$ are used here to denote the five sites
belonging to a pair of neighboring triangular plaquettes as shown in
Fig. 8.

With the help of the Wick's theorem and symmetry arguments the average
of $V$  can  be  reduced  to  the  form
\begin{equation}
\langle V\rangle=\frac{3J_1^2}{4T}(G_4^3-2G_4^2G_3+2G_4G_3^2-G_3^3)
\label{A5} \end{equation}
where
\begin{equation}
G_3  =  g_{01}-\frac{1}{2}g_{13},~~
G_4  =  g_{01}-\frac{1}{2}g_{14}                    \label{A7}
\end{equation}
and
\begin{equation}
g_{ij}\equiv\langle(\varphi_{i}-\varphi_{j})^2\rangle     \label{A8}
\end{equation}
describes the amplitude of flutuations of $\varphi_{i}-\varphi_{j}$
calculated with the help of the harmonic Hamiltonian (\ref{HE}).

The value of $g_{ij}$ for the nearest neighbors  ($g_{01}$) can be
calculated exactly:
\begin{equation}
g_{01}=\frac{T}{J_1}                                     \label{A9}
\end{equation}
whereas numerical calculation of the integrals over Brillouin zone
defining $g_{13}$ and $g_{14}$ gives
\begin{equation}
g_{13}=\left(\frac{3}{2}+\delta\right)\frac{T}{J_1},~~
g_{14}=\left(\frac{3}{2}-\delta\right)\frac{T}{J_1},  \label{A11}
\end{equation}
where $\delta\approx 0.0213$.

Substitution of Eqs. (\ref{A9})-(\ref{A11}) into Eqs.
(\ref{A5})-(\ref{A7}) then gives
\begin{equation}
K_\sigma=\frac{\gamma T^2}{2J_1}                        \label{A12}
\end{equation}
where
\begin{equation}
\gamma=\frac{3}{32}\delta(1+12\delta^2)\approx 2.01\cdot 10^{-3}
                                                        \label{A13}
\end{equation}
which leads to Eq. (\ref{SW}).

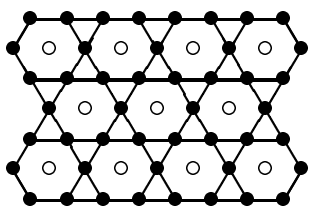
\begin{figure}
%\begin{center}
%\input{fxk-fig1.tex}
%\end{center}
\caption[Fig. 1]
{A {\em kagom\'{e}} lattice (shown by filled circles)
can be constructed by the regular
elimination of one quarter of sites from a triangular lattice.
It consists of triangular and hexagonal plaquettes.}
\end{figure}

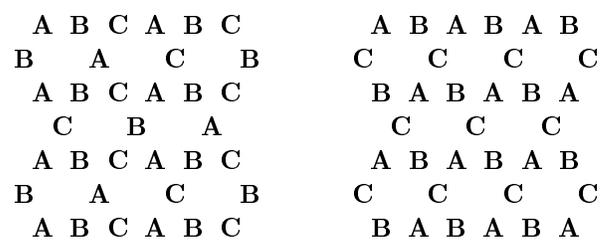
\begin{figure}
%\begin{center}
%\input{fxk-fig2.tex}
%
%(a)\hspace{40mm}(b)}
%
%\end{center}
\caption[Fig. 2]
{The structure of the ground states selected in presence
of interactions with further neighbors.
The letters A, B and C correspond to the three values of
$\varphi_{i}$, which differ from each other by $\pm 2\pi/3$.
(a) a $\sqrt{3}\times\sqrt{3}$ state;
(b) a ${\bf q}=0$ state.}
\end{figure}

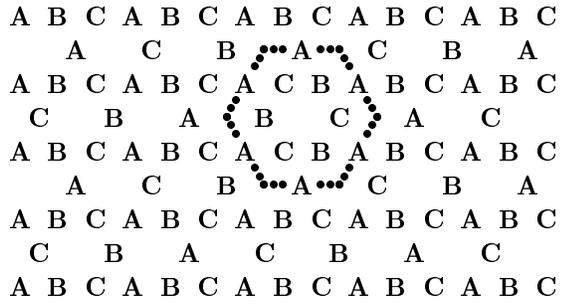
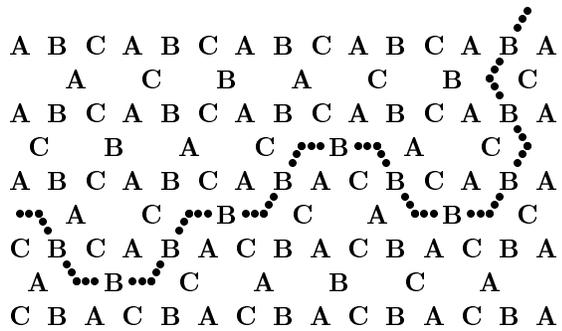
\begin{figure}
%\begin{center}
%\input{fxk-fig3.tex}
%\end{center}
\caption[Fig.3]
{Two examples of a domain wall separating different
$\sqrt{3}\times\sqrt{3}$ states.}
\end{figure}

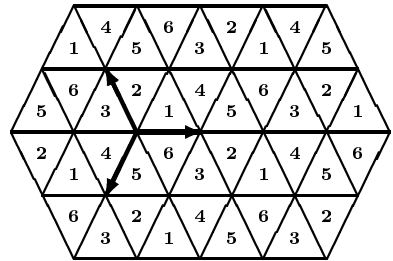
\begin{figure}
%\begin{center}
%\input{fxk-fig4.tex}
%\end{center}
\caption[Fig. 4]
{The triangular lattice ${\cal T}_a$ and its three basic vectors
${\bf a}_\alpha$.
The sites of the dual lattice ${\cal H}_a$ are shown by the numbers
from $1$ to $6$.
The same numbers correspond to physically equivalent states.}
\end{figure}

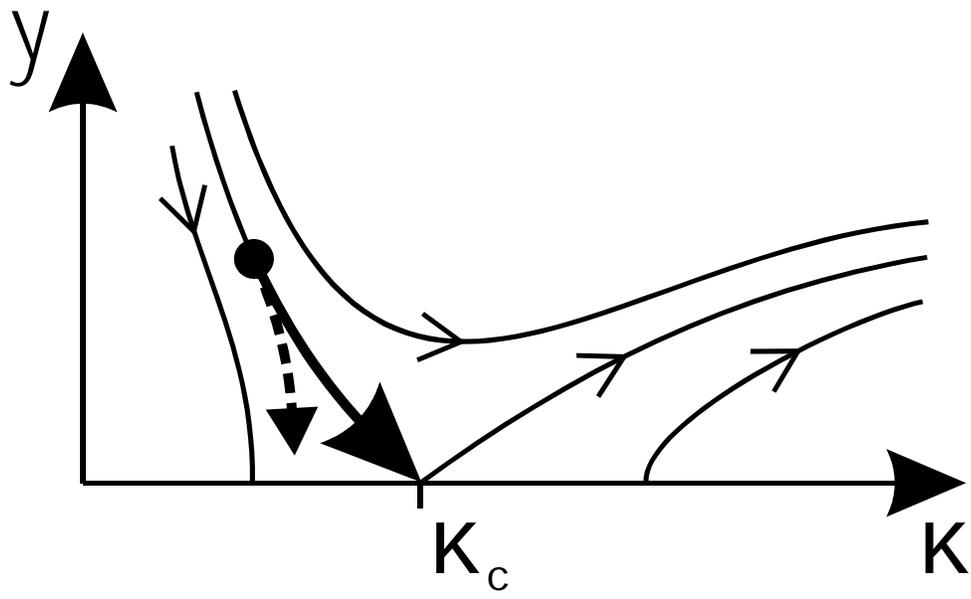
\begin{figure}
%\begin{center}\end{center}
\caption[Fig. 5]
{The schematic flow diagram for Eqs. (8).
The system with only NN interaction and $T=0$ can be associated
with some point (shown by black dot) on the left separatrix.
Dashed arrow shows how the flow is changed at $T>0$ when the
contribution from $z$ has to be taken into account.}
\end{figure}

\begin{figure}
%\begin{center} \input{fxk-fig6.tex} \end{center}
\caption[Fig. 6]
{(a) An example of a fractional vortex.
(b) An example of a dislocation formed by a neutral pair of
fractional vortices.}
\end{figure}

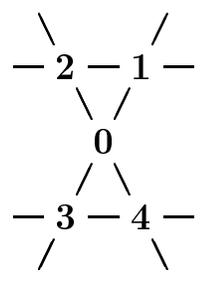
\begin{figure}
%\begin{center} \input{fxk-fig7.tex} \end{center}
\caption[Fig. 7]
{(a) and (b) two examples of a domain wall separating two ${\bf q}=0$
states;
(c) a typical finite size defect on the background of ${\bf q}=0$
state; (d) a kink on a domain wall.}
\end{figure}

\begin{figure}
% \begin{center} \input{fxk-fig8.tex} \end{center}
\caption[Fig. 8]
{The numbering of sites used in the expressions for $H^{(3)}_a$ and
$H^{(3)}_b$.}
\end{figure}

  \end{multicols}
\end{document}